\documentclass[twocolumn,superscriptaddress,prb,floatfix]{revtex4-1}

\usepackage{graphicx}
\usepackage[usenames]{color}
\usepackage{multirow}
\usepackage{verbatim}
\usepackage{amsmath}
\usepackage{amsfonts}
\usepackage{amssymb}
\usepackage{mathrsfs}
\usepackage{url}

\usepackage{float}
\usepackage{setspace,ragged2e}

\begin{document}
\newcommand{\noct}{NiO$_6$}
\newcommand{\nii}{Ni$^{2+}$}
\newcommand{\niii}{Ni$^{3+}$}
\newcommand{\niv}{Ni$^{4+}$}

\newcommand{\dvi}{$d^{6}$}
\newcommand{\ds}{$d^{7}$}
\newcommand{\de}{$d^{8}$}
\newcommand{\del}{$d^{8}$\underline{L}}

\title{High-density electron doping of SmNiO$_3$ from first principles}
\author{Michele~Kotiuga}
\affiliation{Department of Physics and Astronomy, Rutgers, The State University of New Jersey, Piscataway, NJ, USA}
\author{Karin~M.~Rabe}
\affiliation{Department of Physics and Astronomy, Rutgers, The State University of New Jersey, Piscataway, NJ, USA}
\email{karin@physics.rutgers.edu}
\date{\today}
\begin{abstract}
Recent experimental work has realized a new insulating state of samarium nickelate (SmNiO$_3$), accessible in a reversible manner via high-density electron doping. To elucidate this behavior, we use the first-principles density functional theory (DFT) + U method to study the effect of added electrons on the crystal and electronic structure of SmNiO$_3$.  
First, we track the changes in the crystal and electronic structure with added electrons compensated by a uniform positive background charge at concentrations of $\frac{1}{4}$, $\frac{1}{2}$, $\frac{3}{4}$, and 1 electrons per Ni. The change in electron concentration does not rigidly shift the Fermi energy; rather, the added electrons localize on \noct~octahedra causing an on-site Mott transition and a change in the density of states resulting in a large gap between the occupied and unoccupied Ni $e_g$ orbitals at full doping. This evolution of the density of states is essentially unchanged when the added electrons are introduced by doping with interstitial H or Li ions.

\end{abstract}
\maketitle
\section{Introduction}
Doping as a means to control the properties of a material is a well established practice. For conventional semiconductors, the conductivity can be manipulated through low density chemical doping and further altered via the field effect, which is  the basis of operation in many electronic devices~\cite{Li2006}. In transition metal oxides (TMOs) coupling between lattice, charge, spin and orbital degrees of freedom often give rise to many competing phases and interesting properties. Via doping, one can engineer completely new phases, which in some cases possess vastly different properties, such as superconductivity in the cuprates~\cite{Lee2006} and recently, the nickelates~\cite{Li2019}.
More recently high-concentration doping effects have been observed in the interfacial layers of oxide heterostructures~\cite{Ohtomo2004, Mannhart2010} and superlattices~\cite{Chen2013,Chen2013a,Chen2014,Marshall2014,Disa2015}. The short screening length in oxides and the high concentration makes such doped materials useful in nanoscale field-effect devices~\cite{Ahn2003}. Beyond interfacial effects, doping by substitution and interstitials allows one to access these doped phases in bulk, and in some instances, can reach the high densities observed at interfaces.

The well-known rare-earth nickelates, $R$NiO$_3$ ($R$=La-Lu), exhibit a rich temperature-composition phase diagram involving various charge and  magnetic orderings that can be manipulated via doping~\cite{Middey2016,Catalano2018}. Except for LaNiO$_3$, all $R$NiO$_3$ have a metal-to-insulator transition (MIT) accompanied by a symmetry lowering \noct~octahedral breathing distortion of the metallic $Pbmn$ structure resulting in an insulating $P2_1/n$ structure. For $R$=Ce-Nd, an antiferromagnetic (AFM) ordering coincides with the MIT and for $R$=Sm-Lu, the magnetic ordering occurs at a lower temperature. These  materials  are negative charge-transfer compounds in which the O-$2p$ orbitals span the Fermi energy while overlapping the  Ni-$3d$ orbitals, resulting in strongly hybridized Ni-O states and unoccupied oxygen states: 
ligand holes~\cite{Bisogni2016, Haule2017}.  When electron doping the nickelates, these states are paramount in the electron localization and are filled by the added electrons.

Rare-earth nickelates have been electron doped at low concentrations via cation substitution~\cite{Garcia1995,Vobornik1999}, oxygen deficiency~\cite{Nikulin2004, Kawai2009, Shi2013,Dong2017}, and electrolyte gating~\cite{Scherwitzl2010,Shi2013,Dong2017}.
%LaNiO$_3$ and NdNiO$_3$ have been electron doped at low concentrations via cation substitution~\cite{Garcia1995,Vobornik1999},
%and by introducing oxygen vacancies in  NdNiO$_3$~\cite{Nikulin2004, Kawai2009}.
%, which leads to a surprising increase in resistivity~\cite{Nikulin2004, Kawai2009}.
%The rare-earth nickelates have been electron doped at low concentrations via cation substitution~\cite{Garcia1995,Vobornik1999}, and by introducing oxygen vacancies, which leads to an increases in resistivity~\cite{Nikulin2004, Kawai2009, Shi2013, Wang2016}.
%The observed increases in
An increase in resistivity observed with electron doping in rare-earth nickelates and other TMOs has been attributed to the localization of carriers~\cite{Liu2019, Yoo2018}.
%Cation substitution at a concentration of 1-10\% introduces noticeable structural distortions, modifying the MIT. On the other hand, oxygen vacancies in NdNiO$_{3-\delta}$, stable up to $\delta\approx$~0.2, do not modify the transition temperature, but lead to a small increase in the resistivity~\cite{Nikulin2004}. %, but start to destablize around $\delta\approx$0.2 leading to a phase seperation of NiO and Nd$_4$Ni$_3$O$_{10}$.
Electron doping at higher concentrations has been engineered through extreme oxygen deficiency~\cite{Crespin1983,Kawai2009,Sayagues1994,Tung2017,Wang2016,Kotiuga2019}, %LaNiO$_3$, NdNiO$_3$~\cite{Wang2016}, SmNiO$_3$ %, which is superconducting below 15K~\cite{Li2019}, and through 
%Electron doping at higher concentrations has been engineered through
 and through electron rearrangement at the interface within a superlattice~\cite{Chen2013,Chen2013a, XLiu2019}.  %for example those containing LaNiO$_3$~\cite{Chen2013,Chen2013a, XLiu2019}.
Recent experimental work has shown SmNiO$_3$ (SNO) %, a small-gapped semiconductor with a MIT at $\sim$400K,
can be electron doped at extremely high concentrations  via intercalated hydrogen, small alkali and alkaline earth metal ions~\cite{Shi2014,Zhang2018,Sun2018} and the doping can be easily reversed, unlike synthesis-based doping methods. SNO, a narrow-gap semiconductor with a MIT at $\sim$400K, when fully doped becomes % a new insulating
a Mott insulator %phaseemerges
with a gap on the order of 3 eV~\cite{Shi2014}, akin to electron-doped strontium titanate~\cite{Bjaalie2014,Janotti2014}.  Such a phase transition leads us to ask how the added electrons act at the nanoscale resulting in this insulating state. First-principles calculations are uniquely positioned to quantify the changes to the electronic structure and explore how and why the added electrons localize.
%LaNiO$_3$ and NdNiO$_3$ have been electron doped at low concentrations via cation substitution~\cite{Garcia1995,Vobornik1999}, and by introducing oxygen vacancies in  NdNiO$_3$, which leads to a surprising increase in resistivity~\cite{Nikulin2004, Kawai2009, Shi2013, Wang2016}.
%Such an increase in resistivity with electron doping in TMOs has been attributed to the localization of carriers~\cite{Liu2019, Yoo2018}.
%Electron doping at higher concentrations has been engineered through electron rearrangement at the interface within a superlattice, for example those containing LaNiO$_3$~\cite{Chen2013,Chen2013a}.  Recent experimental work has shown SmNiO$_3$ (SNO), a small-gapped semiconductor with a MIT at $\sim$400K, can be electron doped at extremely high concentrations  via intercalated hydrogen, small alkali and alkaline earth metal ions~\cite{Shi2014,Zhang2018,Sun2018} and the doping can be easily reversed, unlike synthesis-based doping methods. When fully doped a new insulating Mott phase emerges with a gap on the order of 3 eV~\cite{Shi2014}, akin to electron-doped strontium titanate~\cite{Bjaalie2014,Janotti2014}.  Such a phase transition leads us to ask how the added electrons act at the nanoscale resulting in this insulating state. First-principles calculations are uniquely positioned to quantify the changes to the electronic structure and explore how and why the added electrons localize.

\begin{table*}[!t]
\caption{Structural and electronic properties of undoped SNO  for various magnetic and charge orderings. For the bond disproportionated structures ($P2_1/n$), two \noct ~octahedral volumes and magnetic moments are given. The non-disproportionated structures relax back to the high symmetry $Pbnm$ space group.}
\centering
\setlength{\tabcolsep}{5pt}
\begin{tabular}{c c c c c c c c c}
  \hline\hline
  Magnetic& a & b & c & $\beta$ & \noct Volume& Total Energy/Ni &Band Gap &Magnetic Moment \\
  Ordering&(\AA)&(\AA)&(\AA)&($^\circ$)&(\AA$^3$)&(eV)&(eV)&($\mu_B$)\\
  \hline\hline
  FM&5.328 &5.553 &7.600&90.031&10.99/9.69&-33.753&0.30&1.49/0.84\\
  AFM(A)& 5.332&5.548&7.594&90.053&11.38/9.35&-33.720&0.77&$\pm$1.61/$\pm$0.27\\
  AFM(G)&5.278 &5.818 &7.421 &90&10.62&-33.648&0.43&$\pm$0.92\\
  AFM(C)&5.270 &5.812 &7.441& 90&10.62&-33.661&0.43&$\pm$1.00\\
  AFM(A)&5.300 &5.724 &7.481&90&10.52&-33.700&0.54&$\pm$1.11\\
\end{tabular}
\label{Table1}
\end{table*}

In this paper, we systematically investigate the effect of electron doping on room temperature SNO using first-principles calculations. 
We study the effect of added electrons on the structure, density of states, orbital occupation and magnetic moments of the nickel. Furthermore, we compare electron doping with added electrons and a positive compensating background charge to the inclusion of interstitial hydrogen and lithium to study the effect of the doping mechanism and separate the effect of the added electron from that of the doping mechanism. The added electrons do not introduce typical defect-like states~\cite{Freysoldt2014} or a rigid shift of the Fermi energy, but rather localize on NiO$_6$ octahedra, moving an unoccupied state to the valence band. At a doping concentration of 1 added $e^-$/Ni, a novel insulating Mott phase of SNO is realized.
These localized electrons open up a new class of materials and phases.

\section{Methods}
Our first-principles DFT + U calculations are carried out using the Perdew-Burke-Enzerhof (PBE) functional~\cite{Perdew1996,Perdew1997} as implemented in VASP~\cite{Kresse1996,Kresse1999} with an energy cutoff of 520 eV and the Sm\_3, Ni\_pv, O  projector augmented wave (PAW)~\cite{Blochl1994paw} potentials provided with the VASP package.  The Sm\_3 PAW potential includes the $f$-electrons in the core. When treating interstitial hydrogen or lithium, we use the H or Li\_sv potentials, respectively.  In both DFT+U and dynamical mean field theory (DMFT) calculations of nickelates ~\cite{Park2012,Park2014,Park2014b,Subedi2015,Varignon2017,Mercy2017,Hampel2017,Hampel2019}, large values of U are used to reproduce thermodynamic quantities ~\cite{Wang2006,Lee2009}, intermediate values to reproduce insulating states, and small values to reproduce the low temperature magnetic properties~\cite{Varignon2017,Mercy2017, Hampel2017, Hampel2019}. We include a Hubbard U (within the rotationally invariant method of Liechtenstein et al.~\cite{Liechtenstein1995}) with U=4.6 eV and J=0.6 eV. We choose this intermediate value as we are not focusing on low temperature or magnetic properties. Using a smaller U (U=2.6 eV), there is no qualitative change to the results 
 (see SI sec. B).  To accommodate the $a^-a^-c^+$ tilt pattern present for all phases of SNO, we used a 20 atom  $\sqrt{2}\times\sqrt{2}\times2$ supercell containing four Ni sites. To explore monoclinic phases a $\beta=90.75^{\circ}$ was introduced in the starting structures.   Structural relaxations of both the internal coordinates and the lattice parameters were carried out with Gaussian smearing with $\sigma=0.1$ eV and a Monkhorst-Pack $k$-mesh of 6$\times$6$\times$4 such that the forces are less than 0.005 eV/\AA.  The density of states calculations were performed using  the tetrahedral method with Bl\"{o}chl corrections~\cite{Blochl1994}. The projected density of states (PDOS) plots were generated using a $\Gamma$-centered $k$-mesh and the site projected scheme of pymatgen~\cite{Ong2013}.

To study the effects of high-density electron doping, we increase the number of electrons while adding a uniform positive compensating background charge, or ``jellium"~\footnote{This achieved by increasing the flag NELECT in our VASP calculations.}. By using this method to model doping, we separate the effects stemming from the added electrons from the effects due to specifics of the doping mechanism. We explore the changes to electronic structure, Ni magnetic moments and local structural changes. 
When using a uniform positive compensating background charge, which is unphysical in itself, an unphysically large expansion of the total volume (25\% at a concentration of 1 added e$^-$/Ni) occurs. We, therefore, keep the total volume fixed while only relaxing the internal coordinates.

We added one electron to a $\sqrt{2}\times\sqrt{2}\times2$ supercell of undoped SNO, a doping concentration of $\frac{1}{4}$ e$^-$/Ni, relaxing only the internal coordinates. 
To clearly track the changes as the doping concentration increases, we used this relaxed structure as the starting structure for a doping concentration of $\frac{1}{2}$ e$^-$/Ni. Thus, we added one electron per supercell to this structure and, again, relaxed only the internal coordinates.  We continued in this manner until we reached a doping concentration of 4 electrons per supercell, or 1 e$^-$/Ni. For doping concentrations of $\frac{1}{4}$ and 1 e$^-$/Ni, we compare doping with a jellium background to interstitial hydrogen and lithium ions.

To find the low energy structure for added H atoms at a concentration of $\frac{1}{4}$ H/Ni, we constructed 
a series of $\sqrt{2}\times\sqrt{2}\times2$ supercells of undoped SNO  with an added H atom placed in the plane bisecting the Ni-O-Ni bonds at the O for both an apical and basal sites. The H atoms were placed approximately $\pm$1\AA~ away from the O along the crystallographic axes lying in the plane. 
For added Li, we followed the same procedure placing the Li atoms approximately $\pm$2\AA~away from the the apical or basal O. For the higher concentrations of added atoms, we followed a prescription analogous to that for added electrons: using the relaxed structure at a given concentration as the starting structure for the next highest concentration. We used the lowest energy position of the added ion from the $\frac{1}{4}$ doping case to determine where to place the additional ions~\cite{Zhang2018,Sun2018}. We performed relaxations of the internal coordinates with fixed volume, constrained to (110) epitaxial strain, as well as a full volume relaxation.

\section {Results and Discussion}
\subsection{Pristine SmNiO$_3$}
\begin{figure}[t]\vspace{-0pt}
\center
\includegraphics[width=0.40\paperwidth]{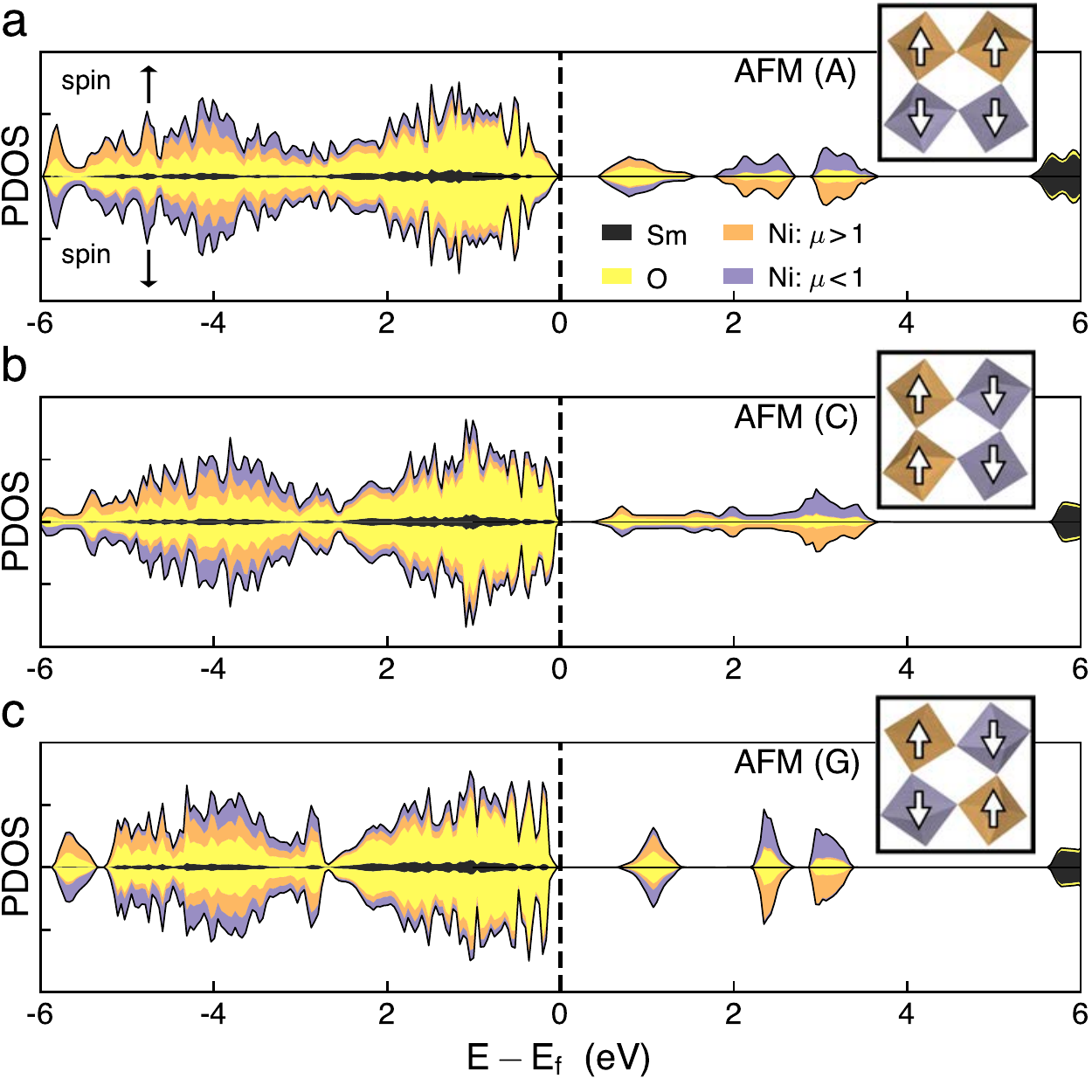}
\caption{Projected density of states (PDOS) for undoped SNO for the non-disproportionated cases in Table \ref{Table1}, which are dominated by O (yellow) and Ni (orange for a positive magnetic moment and purple for a negative moment), while Sm (dark gray) has a minimal contribution. Insets: the four \noct~octahedra supercell with the Sm and O atoms omitted for clarity. The arrows schematically denote the spin of the occupied Ni $e_g$ levels.}
\label{fig:SNO}
\end{figure}

Results for the properties of undoped SNO are reported in Table~\ref{Table1}. The lowest energy structure exhibits a bond disproportionation ($P2_1/n$), consistent with previous calculations~\cite{Varignon2017, Liu2019} ( see SI sec. A for electronic structure).
As room temperature SNO does not display a strong bond-disproportionation signature~\cite{Alonso1999,Medarde2009},  we focus on metastable non-disproportionated structures. 
Initializing the magnitude of all Ni moments to be the same value of 5$\mu_B$, we find that the structure relaxes back to a non-disproportionated orthorhombic ($Pbnm$) structure. We find three stable magnetic orderings: 
layered (A), columnar (C), and rock-salt (G), all of which share the same key features of their electronic structures (see Fig.~\ref{fig:SNO}). Both the occupied and unoccupied states within the vicinity of the Fermi level are dominated by O-$2p$ and Ni-$3d$ states with minimal  Sm character. The large hybridization along with the O-$2p$ character in the unoccupied states is  consistent with the electronic structure of a negative-charge transfer compound. 
The electronic structure has a more dispersive unoccupied states when neighboring Ni sites have aligned spins, Fig.~\ref{fig:SNO} (a) \& (b), stemming from an increased overlap of these states. 
The unoccupied DOS is comprised of Ni $e_g$ states hybridized with O-$2p$ states. There are three unoccupied states per Ni site, two in one spin channel and one in the other. 
All Ni have a magnetic moment $\sim$1$\mu_B$, suggesting the presence of low-spin \niii. Considering the hybridized states as ones associated with the entire \noct~octahedron, it is more appropriate to write the electronic configuration as \del, i.e. \de~with an oxygen ligand-hole. Schematically, we view this as one occupied Ni $e_g$ state (Fig.~\ref{fig:SNO} insets)

\subsection{Electron doped SmNiO$_3$: Jellium Background}
\begin{figure}[!t]
\center
\includegraphics[width=0.40\paperwidth]{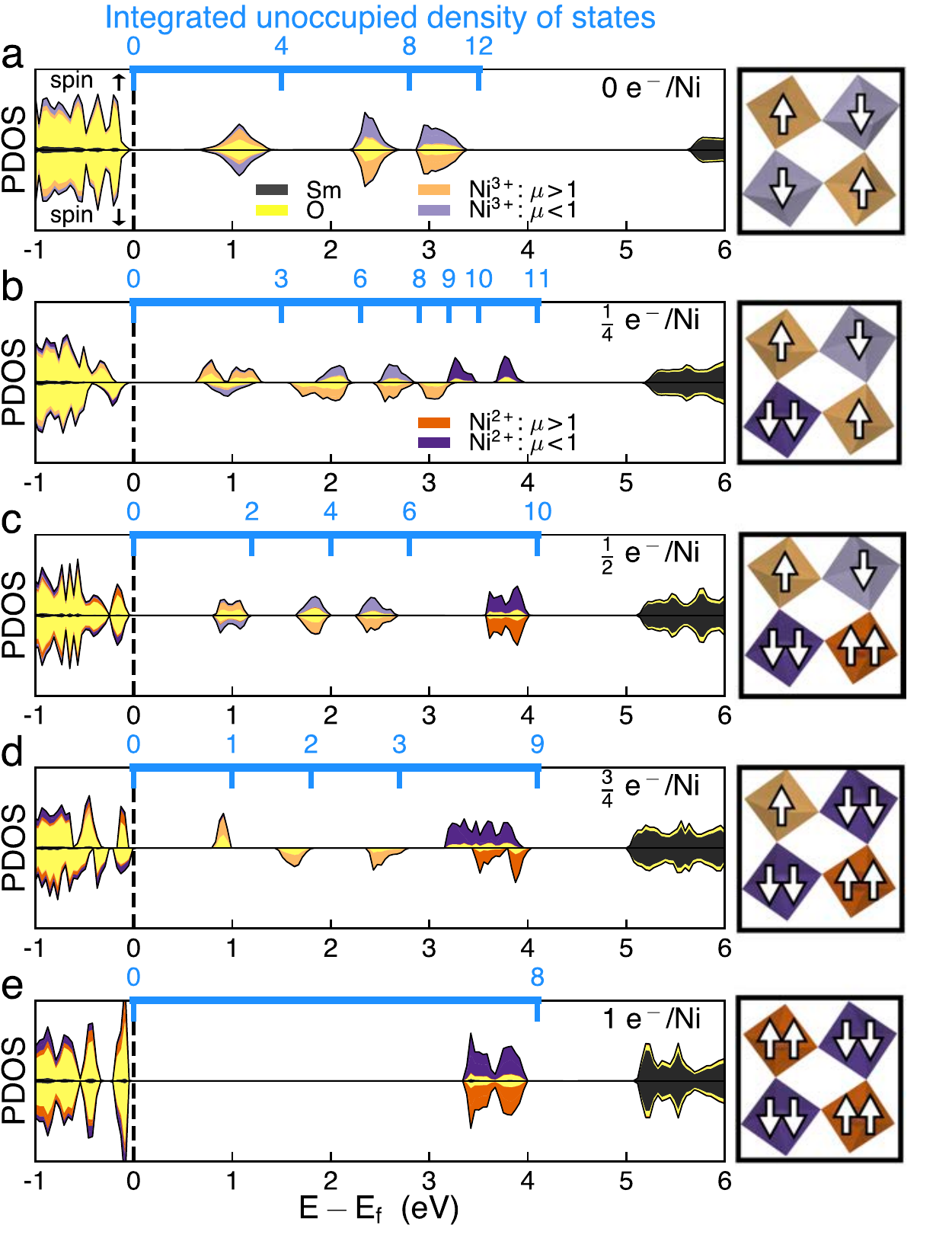}
\caption{(a)-(e) Spin-polarized PDOS for electron-doped SNO and schematic of Ni $e_g$ occupation at concentrations of 0, $\frac{1}{4}$, $\frac{1}{2}$, $\frac{3}{4}$, and 1 electron/Ni for an AFM G-type magnetic ordering. The integrated unoccupied DOS is displayed along the top axis for $E-E_f\in[0,4.5]$. The element specific PDOS is shown in dark gray for Sm, yellow for O, orange and purple for Ni.  \niii~with a positive moment is shown in light orange and with a negative moment in light purple and \nii~with a positive moment is shown in dark orange and and with a negative moment in dark purple. The same color scheme is used for the octahedra in the panels to the right where the arrows schematically denote the spin of the occupied Ni $e_g$ levels.}
\label{fig:dDOS}
\end{figure}

\begin{figure}[!t]
\center
\includegraphics[width=0.40\paperwidth]{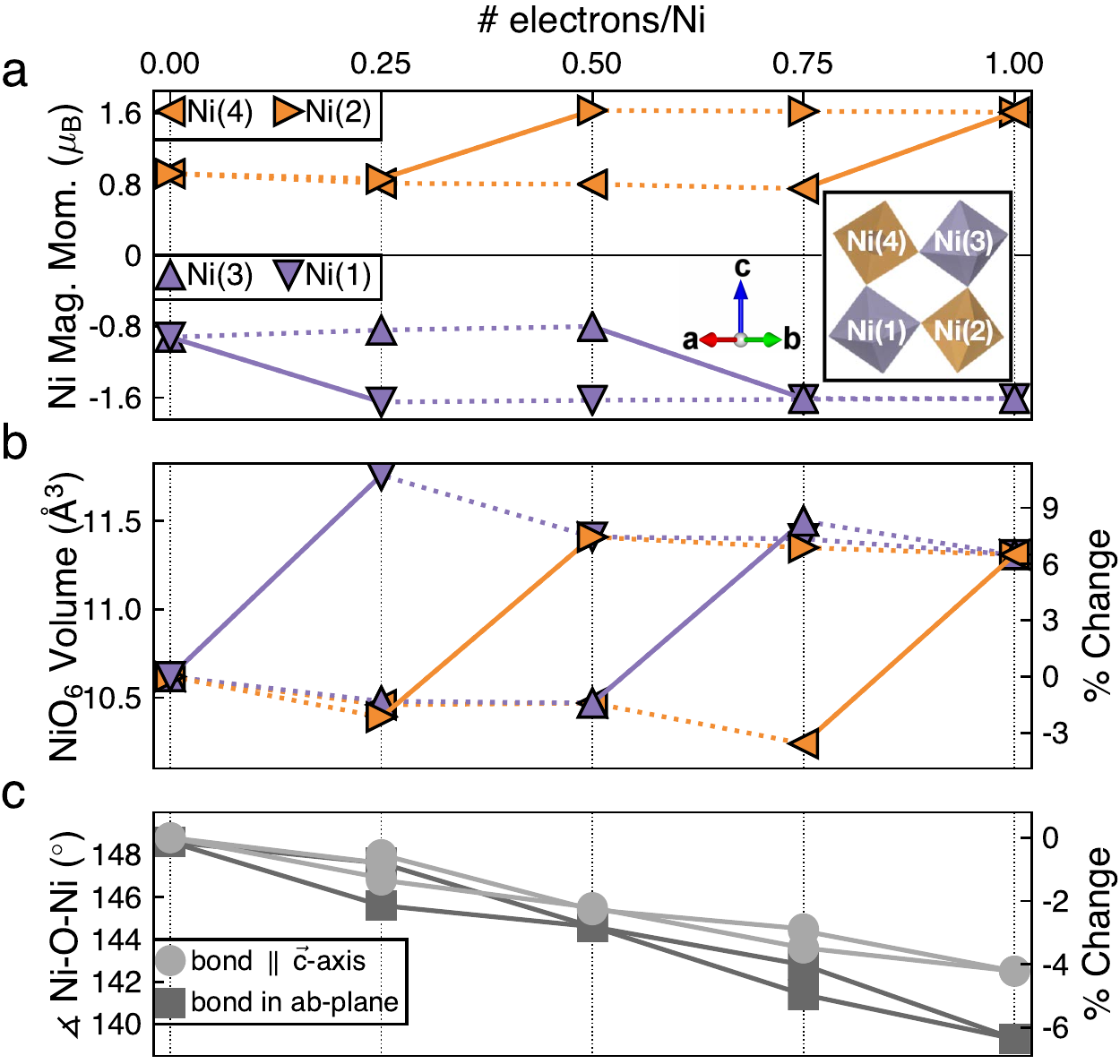}
\caption{The change in (a) the magnetic moment and (b)the \noct~octahedral volume (b) for each Ni site, with orange and purple denoting the Ni sites with a positive and negative moment, respectively. Inset: the calculation supercell relative to the crystallographic axes with each Ni numbered. (c) The change in the Ni-O-Ni angles with electron doping. 
  The bonds parallel to the $\vec{c}$-axis are in light gray circles and those parallel to the $\vec{ab}$-plane are in dark gray squares. }
\label{fig:dProp}
\end{figure}

The PDOS of electron doped SNO is reported in Fig.~\ref{fig:dDOS}; and the corresponding changes to local Ni moments and the structural changes are summarized in Fig.~\ref{fig:dProp}. Adding one electron to undoped SNO with G-type magnetic ordering (i.e. a doping concentration of $\frac{1}{4}$ e$^-$/Ni) results in dramatic changes as the added electron localizes on one \noct~octahedron. 
Without a loss of generality, we break symmetry and localize the electron 
on the \noct~octahedron containing Ni(1) (inset Fig.~\ref{fig:dProp}). As shown in Fig.~\ref{fig:dDOS}(b), 
the integrated unoccupied density of states of the feature closest to the Fermi energy decreases from 4 to 3: now with one unoccupied state associated with each of the unaffected Ni sites: Ni(2)-(4).

The spin-down Ni-$3d$/O-$2p$ state associated with Ni(1) occupied by the added electron moves to the top of the valence band and is dominated by oxygen character essentailly filling the ligand hole. 
The two remaining unoccupied Ni(1) $3d$-states are spin-up $e_g$ bands. These states, in dark purple, are pushed up in energy from $\sim$3eV to $\sim$3.5eV above the Fermi energy and show a noticeable decrease of hybridization with oxygen.   The splitting of the features of the PDOS from Fig.~\ref{fig:dDOS}(a) to (b) results from local symmetry breaking in the distorted structure. Due to Hund's coupling,  \nii~is in a high-spin $S=1$ configuration and the magnitude of the magnetic moment of Ni(1) increases from 0.92$\mu_B$ to 1.65$\mu_B$, while all moments on the other nickels remain relatively unchanged decreasing from  0.92$\mu_B$ to 0.81-0.86$\mu_B$ (Fig.~\ref{fig:dProp}(a)). The volume of the \noct~octahedron containing Ni(1) increases by $\sim$12\%. To accommodate this expansion, all octahedra tilt more while the volumes of the three other octahedra slightly decrease  as the total unit-cell volume is kept constant (Fig.~\ref{fig:dProp}(b) \& (c)). 

\begin{figure}[!t]
  \center
  \includegraphics[width=0.40\paperwidth]{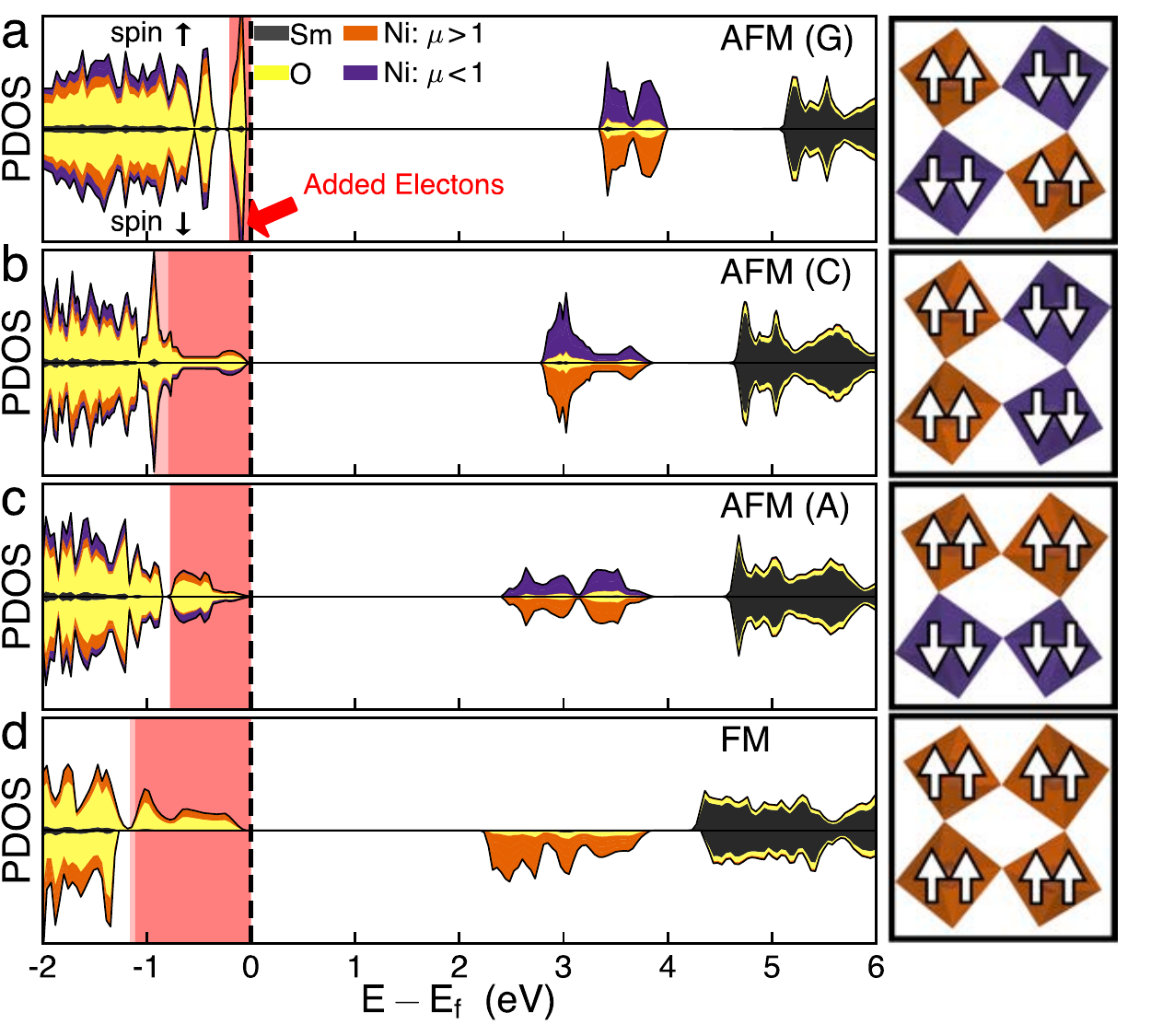}
  \caption{Spin-polarized PDOS for electron-doped SNO and schematic of Ni $e_g$ occupation at a concentrations of 1 electron/Ni for various magnetic orderings: (a) AFM G-type,  (b) AFM C-type, (c) AFM A-type, (d) FM. The portion of the valence band associated with the added electrons is shaded in red, which is dominated by oxygen for all magnetic orderings. The states associated with the added electrons are separate from the original valence band for AFM-G and AFM-A orderings. For AFM-C and FM orderings, these state partially overlap with the top of the original valence band. Where they overlap they are shaded in a lighter red. The element specific PDOS is shown in dark gray for Sm, yellow for O, orange for Ni with a positive moment and purple for Ni with a negative moment. The arrows schematically denote the spin of the occupied Ni $e_g$ levels.}
  \label{fig:eSNO}
\end{figure}

For two added electrons we find that the lowest energy configuration has zero net magnetization, thus the second added electron localizes on Ni(2), which has a positive magnetic moment resulting in no net magnetization of the cell while restoring some symmetry. The added electron occupies the spin-up Ni(2)-3$d$/O-2$p$ state, moving it to the top of the valence band and, again, decreases the integrated unoccupied density of states by one. The new state in the valence band is again dominated by oxygen character while the remaining unoccupied (spin-down) Ni(2) states are pushed away from the Fermi energy and show a marked decrease in the oxygen hybridization (Fig.~\ref{fig:dDOS}(c)). The magnetic moment of Ni(2) increases to 1.63$\mu_B$ and the moment of  Ni(1) decreases slightly to match. The moments of Ni(3) \&\ (4) again decrease very slightly to 0.80$\mu_B$.
The two octahedra containing localized electrons are now the same so the volume of the octahedron containing Ni(1) decreases slightly and the volume of the octahedron containing Ni(2) increases such that both are $\sim$7\%  larger that those of undoped SNO. All the octahedra tilt more while the volumes of the remaining octahedra slightly decrease. 
The third electron localizes on Ni(3), again breaking symmetry, 
leaving only one unoccupied state within 1 eV of the Fermi energy as yet another oxygen dominated state is moved to the top of the valence band. The remaining unoccupied states associated with Ni(3) are pushed up in energy (Fig.~\ref{fig:dDOS}(d)).  At the same time, the magnetic moment of Ni(3) increases to 1.61$\mu_B$, while the moments of Ni(1) \& Ni(2) decrease to match. The moment of Ni(4) decreases to 0.75$\mu_B$ as its octahedron is quite compressed by the other \noct~octahedra, which have all expanded.

Upon adding the fourth electron, localizing on Ni(4), we reach a doping concentration of 1 e$^-$/Ni. The last remaining unoccupied state within 1 eV of the Fermi energy moves to the top of the valence band and the remaining states associated with Ni(4) are pushed up in energy.  The remaining unoccupied Ni states have been pushed up in energy and are now $\sim$3.5 eV above the Fermi energy with negligible oxygen hybridization.
Furthermore,  all nickels have a similar magnetic moment and are situated in a similarly sized \noct~octahedron.

\begin{figure*}[t]
\center
\includegraphics[width=.82\paperwidth]{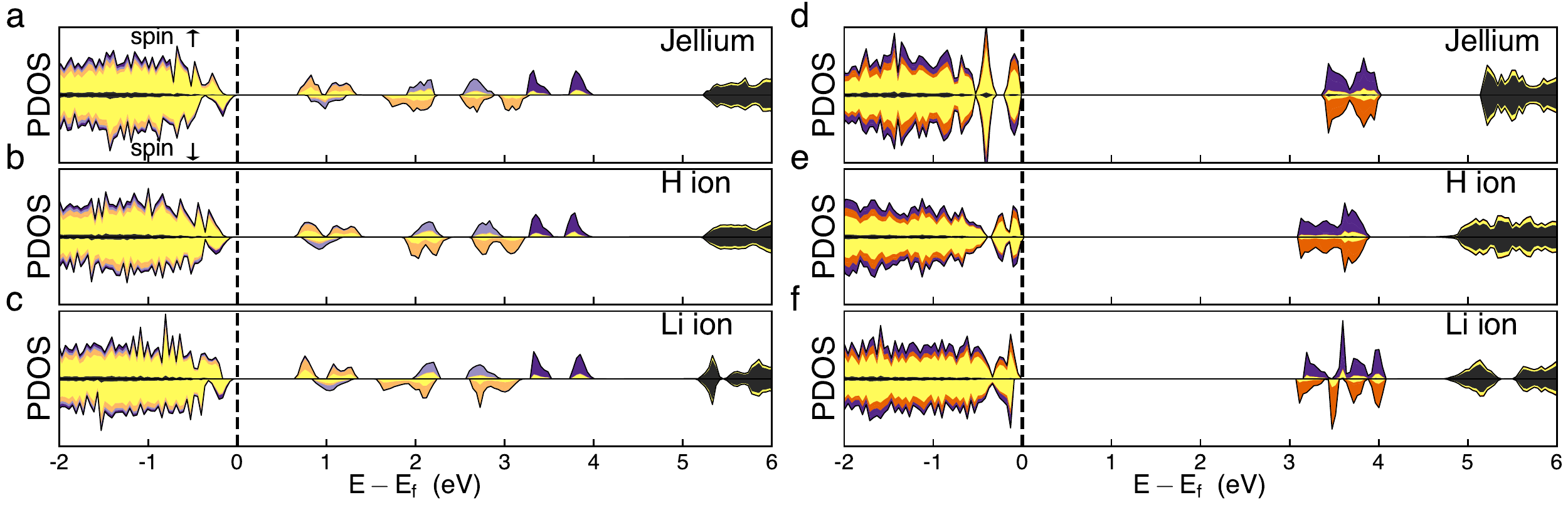}
\caption{Spin-polarized PDOS for electron-doped SNO with (a) 1/4 added e$^-$/Ni with a positive compensating background charge; (b) 1/4  H/Ni; (c) 1/4  Li/Ni;  (d) 1 added e$^-$/Ni with a positive compensating background charge; (e) 1  H/Ni; and, (f) 1 Li/Ni. The element specific PDOS is shown in dark gray for Sm, yellow for O, orange and purple for the Ni with a positive and negative moment, respectively.  \niii~with a positive moment is shown in light orange and with a negative moment in light purple and \nii~with a positive moment is shown in dark orange and and with a negative moment in dark purple.}
\label{fig:inter}
\end{figure*}

 Although our calculations use the AFM (G) case as a starting point, we find similar results for electron doping the AFM (A) or (C) cases: for a given electron doping concentration, we find the lowest energy configuration has the lowest net magnetization. The localized electrons result in high-spin Ni$^{2+}$, consistent with Hund's rules. In fact we find the low-spin configuration is $\sim$0.5 eV/Ni higher in energy. At a doping concentration of 1 $e^-$/Ni, we find the energy difference from the top of the unoccupied Ni states to the top of the valence band is insensitive to magnetic order (see Fig.~\ref{fig:eSNO}). The band width of the unoccupied Ni states increases when neighboring Ni sites have aligned spins, as seen in undoped SNO, resulting in the smallest band gap for FM ordering. The top of valence band, comprised of unoccupied states of undoped SNO now occupied by added electrons, is chiefly of oxygen character: the filled ligand holes. Minimal oxygen character is found in the unoccupied DOS: all Ni sites now have an \de~configuration with no oxygen ligand holes (See SI sec. C for band structures).

 Even though the band gap remains relatively constant for a doping concentration of less than 1 e$^-$/Ni, the integrated density of states of the unoccupied states decreases as the doping concentration increases, consistent with the experimentally observed increase in resistivity as a function of doping. Ultimately, the experimentally observed resistivity saturates at a concentration of 1 e$^-$/Ni, again consistent with the radical increase in the band gap. Thus, by modeling electron doping with a positive compensating jellium background, we have successfully captured the the changes in the electronic structure and local properties of the \noct~octahedra. To better understand why this works, in the next section we compare these results with a jellium background to results with explicitly added ions.

\subsection{Electron doped SmNiO$_3$: Interstitial Ions}

To explore the effect of explicitly including ions in our calculations, we electron dope SNO by adding a neutral hydrogen or lithium. In both cases, the valence electron of the added ion localizes on a nearby NiO$_6$ octahedron resulting in a $\sim$10\% expansion of the octahedron similar to the jellium case.
The lowest energy configuration for both ions is to reside in the tetrahedral space formed between two NiO$_6$ octahedra that have a relative out-of-phase tilt while canting toward one another.
 In the case of hydrogen, the H$^+$ ion sits off center and binds to a basal oxygen (see SI Fig. 9)~\cite{Zhang2018}. For lithium, the Li$^+$ ion is centered and is tetrahedrally coordinated by oxygen (see SI Fig. 10)~\cite{Sun2018}.
 At a concentration of $\frac{1}{4}$  e$^-$/Ni, one  Ni-$3d$/O-$2p$ state is occupied and moved to the top of the valence band, as we found in the jellium case (see Fig.~\ref{fig:inter}(a-c)). Furthermore, the localized electron increases the magnetic moment of one Ni to 1.65$\mu_B$ for an added H or 1.67$\mu_B$ for an added Li, leaving the other moments relatively unchanged. For higher concentrations, the total magnetization is minimized as is observed in the jellium case. At a concentration of 1 e$^-$/Ni a large gap is observed similar to the jellium case (see Fig.~\ref{fig:inter}(d-f)). Unlike the jellium case, the expansion changes the shape of the octahedron and prioritizes creating a binding environment for the ion. The only difference in the electronic structure arises from the local distortion of the NiO$_6$ octahedra due to the added ions causing a splitting of the unoccupied Ni $e_g$ states. Finally, we allow the volume to relax the epitaxial constraint of the experimental thin films, namely an expansion of the (110) orthorhombic direction. We find a volume increase of $\sim$6\% and $\sim$11\% for added H and Li, respectively, in good agreement with experiment~\cite{Zhang2018,Sun2018}. This expansion does not impact the qualitative nature of the electronic structure, leading only to a small decrease in the band gap (see SI Fig. 11). 

 This analysis shows that for doping by intercalation the changes to the magnetic properties and electronic structure are captured by adding electrons with a uniform jellium background charge: it is only necessary to explicitly include the interstitial ions to study structural changes. This positions jellium background calculations to help us in predicting new phases of materials enabled by high-density doping, % due to charge localization.
 not only via electrolyte gating but via doping by intercalation and oxygen deficiency.
 New phases stemming from carrier localization have% This positions jellium background calculations to help us in predicting new phases of materials via high-density doping due to charge localization. This high-concentration doping mechanism has
 already been realized in other $R$NiO$_3$~\cite{Sun2018}, and is expected to occur in other negative charge-transfer compounds where unoccupied ligand states would play a part in the electron localization. 
 
\section{Conclusion}
  In summary, electron doping of SNO results in dramatic changes to the electronic structure. 
  These changes to the electronic structure cannot be understood by rigidly shifting the Fermi energy, but stem from electron localization causing an on-site Mott transition. 
  At a doping concentration of one electron per Ni, all \niii~are converted to \nii~resulting in a Mott gap between occupied Ni $e_g$ states hybridized with O and unoccupied Ni $e_g$ states of the opposite spin with negligible O hybridization. 
  This behavior is captured in first-principles calculations even when the added electrons are accompanied by a jellium background, enabling the identification of new systems with these electron-doping-induced effects.

\begin{acknowledgments}
We thank Cyrus E. Dreyer and Shriram Ramanathan for useful discussions and acknowledge financial support from Office of Naval Research Grant N00014-17-1-2770.
\end{acknowledgments}

\bibliographystyle{apsrev4-1}
\bibliography{SNOwSI}

\onecolumngrid

\appendix
\newpage
\centering{\bf Supplemental Materials: High-density electron doping of SmNiO$_3$ from first principles}

\section{Bond-Disproportionated density of states}
\begin{figure}[!h]
\center
\includegraphics[width=0.5\textwidth]{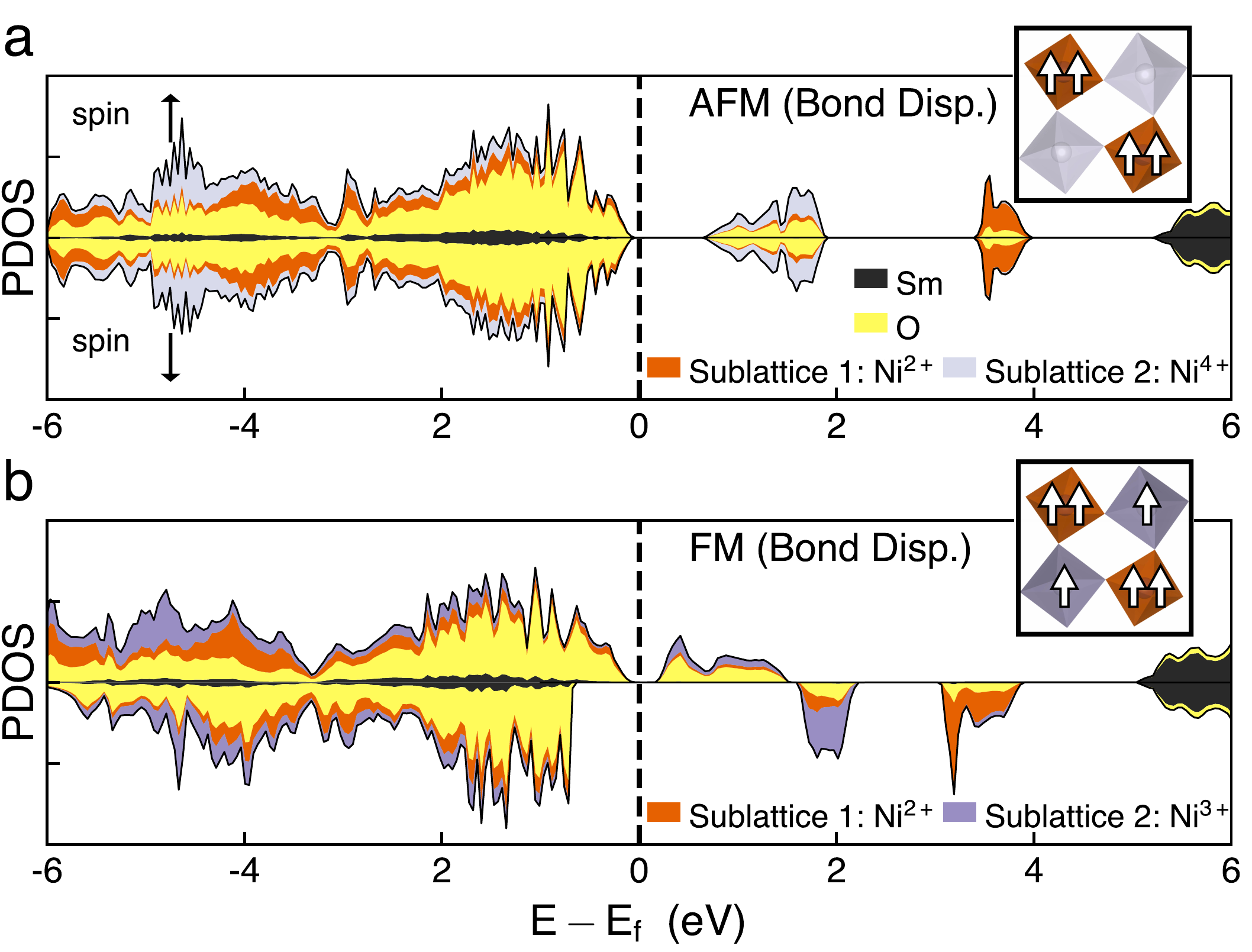}
\caption[S1]{Projected density of states (PDOS) for bond-disproportionated structures in Table I of the main text, which is dominated by O (yellow) and Ni (orange and purple), while Sm (dark gray) has a minimal contribution. Insets: the calculation supercells with the Sm and O atoms omitted for clarity. The two colors distinguish the two Ni sublattices in the monoclinic structure, while the lightest hue accompanies octahedra containing \niv, the medium \niii~and the darkest \nii. The arrows schematically denote the Ni $e_g$ occupation.}
\label{fig:BD}
\end{figure}
\flushleft
As mentioned in the main text, in general the electronic structure has more dispersive states when neighboring Ni sites have aligned spins most likely due to an increased overlap of these states. The bond disproportionation results in different volumes of the \noct~octahedra of each sublattice. The sublattice with larger octahedra,  sublattice 1 in Fig.~\ref{fig:BD}, also has a larger magnetic moment. For the AFM bond disproportionated result we observe a qualitatively similar occupied DOS to that of the non-bond-disproportionated results.  The unoccupied DOS, on the other hand, is quite different with only two lobes. The lower one, associated with sublattice 2, has 2 bands per spin channel per Ni, i.e. \niv(\dvi) with no $e_g$ occupation; however, the high O character of this lobe again suggests the presence oxygen ligand holes. On the other hand, the upper lobe, associated with sublattice 1, is spin polarized, with 2 spin up or down bands on each Ni. Together with the negligible O contribution, this implies a \nii(\de) state with no ligand holes.
\\
The FM bond-disproportionated case has a noticeably different electronic structure. The occupied hybridized Ni-O states are reminiscent of the other scenarios; however, due to the FM ordering, the integrated DOS of the spin minority channel (here, spin down) is 1 state/Ni less than the majority channel. The majority spin channel has four very dispersive unoccupied states near the Fermi energy comprised of Ni-O hybridized states associated with sublattice 2. In the minority spin channel, a lobe associated with each sublattice is situated higher in energy both displaying a reduction if the hybridization with O. Like the AFM ordered bond-disproportionated case, this suggests the Ni of sublattice 1 have a \de configuration with no oxygen ligand holes and as well as the presence of oxygen ligand holes on sublattice 2.
\newpage
\section{Density of states with electron doping using smaller U}
\begin{figure}[!h]
\center
\includegraphics[width=0.5\textwidth]{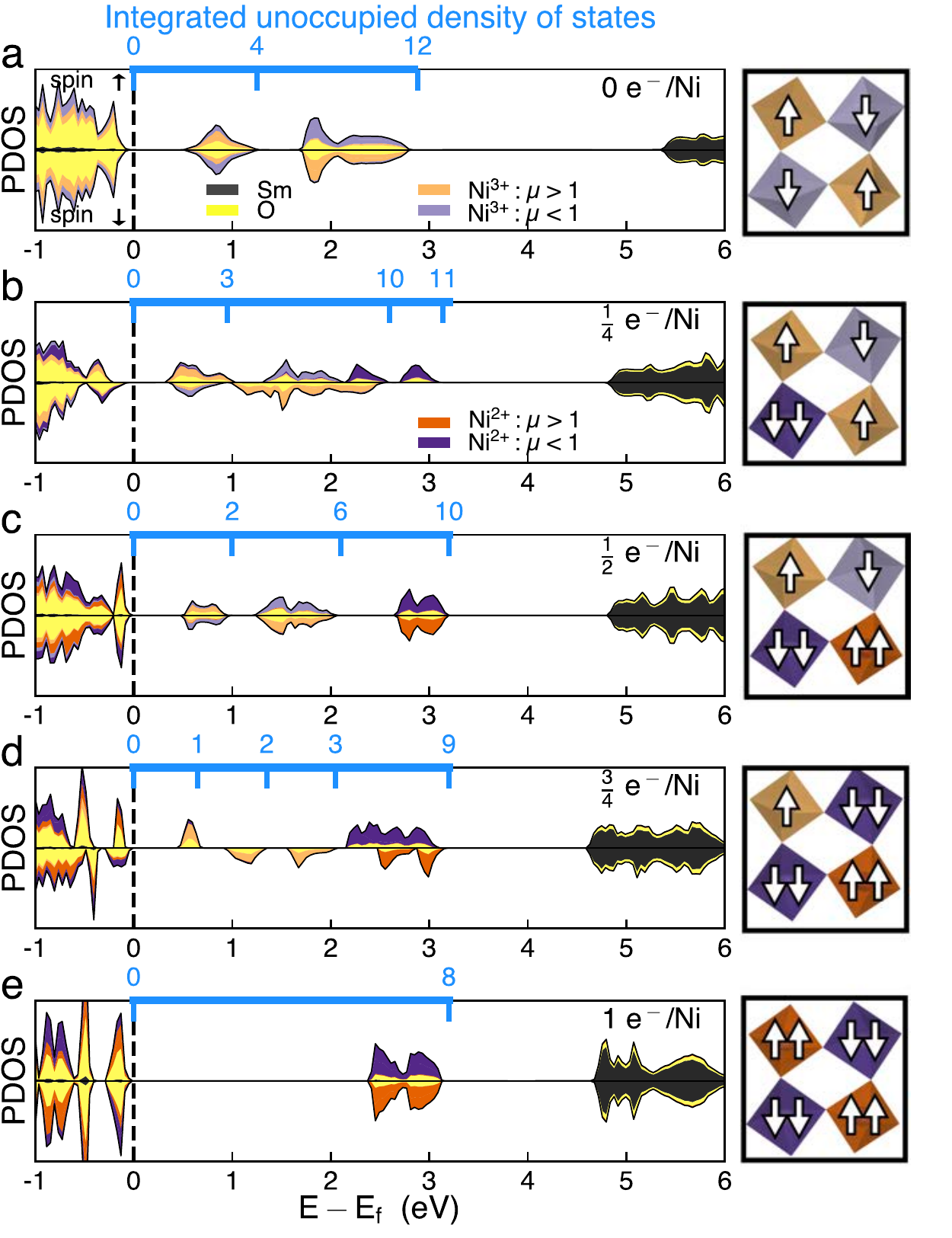}
\caption{Spin-polarized PDOS for electron-doped SNO and schematic of Ni $e_g$ occupation at concentrations of 0, $\frac{1}{4}$, $\frac{1}{2}$,$\frac{3}{4}$, and 1 electron/Ni in panels (a)-(e) for an AFM G-type magnetic ordering. The element specific PDOS is shown in yellow for O, orange and purple for Ni with a positive or negative magnetic moment, respectively, and dark gray for Sm. \niii~is displayed in the lighter hues and \nii~in the darker hues. The integrated unoccupied DOS is displayed along the top axis for $E-E_f\in[0,4.5]$. The arrows schematically denote the Ni $e_g$ occupation}
\label{fig:U2}
\end{figure}

We find the same evolution of the density of states with electron doping as in the main text, here, using a smaller value of U: U=2.6 eV, J=0.6 eV, see  Fig.~\ref{fig:U2}. As we are using a smaller value of U, the band gap is smaller that the results reported in the main text using U=4.6 eV, J=0.6 eV.
\newpage
\section{Band structure of doped SNO}
\begin{figure}[!h]
\center
\includegraphics[width=0.99\textwidth]{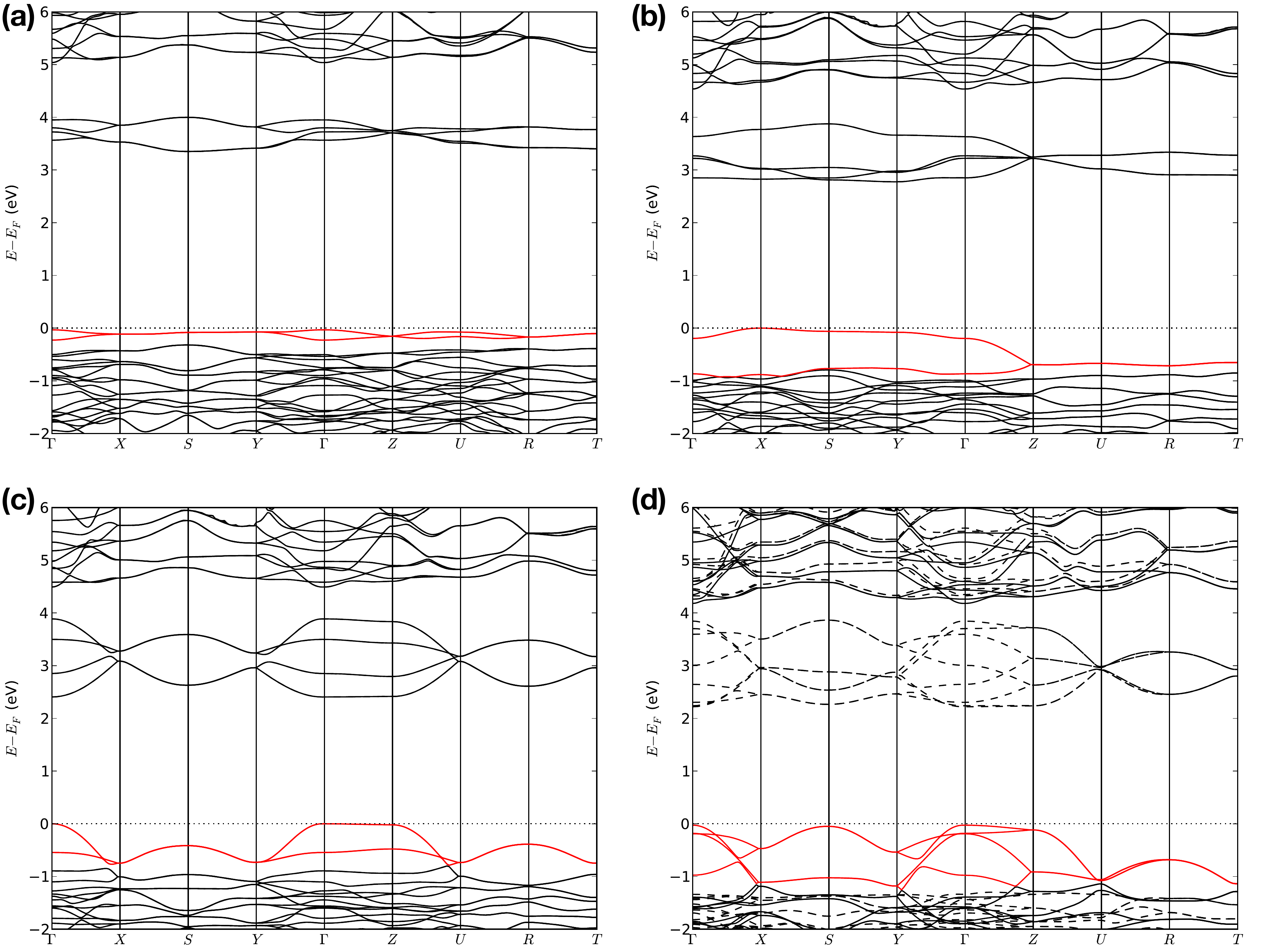}
\caption{Band structure of SNO doped with 1 e$^-$/Ni with (a) G-, (b) C-, and (c) A-type antiferromagnetic orderings and (d) ferromagnetic ordering. The bands in red are the new bands from the added electrons. For the ferromagnetic case the spin up bands are solid and the spin down bands are dashed. As the antiferromagnetic cases are not spin polarized, the spin up and spin down bands lie on top of one another.}
\label{fig:band}
\end{figure}
\newpage
\section{Explicit (added interstitials) electron doping}
\begin{figure}[!h]
\center
\includegraphics[width=\textwidth]{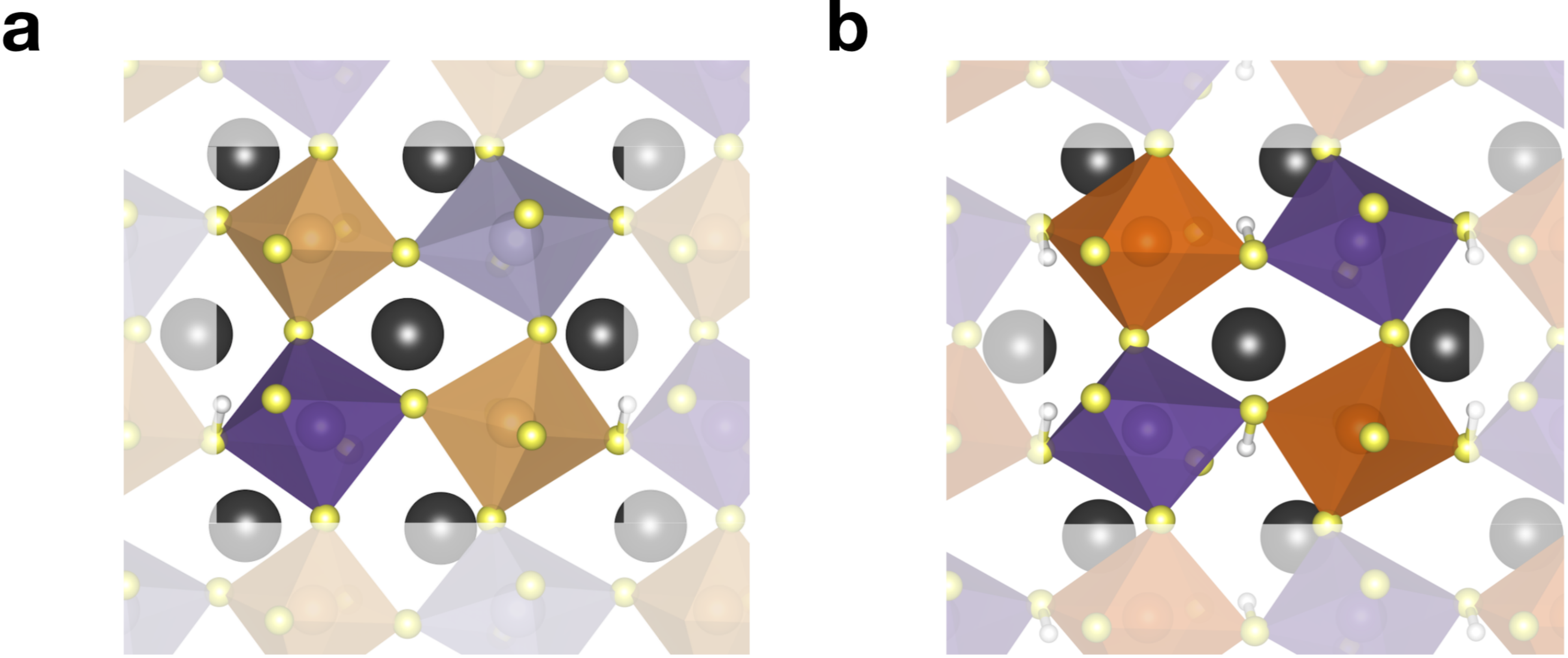}
\caption{One $\sqrt{2}\times\sqrt{2}\times 2$ cell for the (a) low energy structure for H$_{\frac{1}{4}}$SNO; and, (b) the low energy structure for HSNO. Sm shown in dark gray; Ni in purple and orange, where the lighter hues indicate Ni$^{3+}$ and the darker hues Ni$^{2+}$; O in yellow; and, H in white. The hydrogen bind to an oxygen with a bond length of $\sim$ 1\AA in the tetrahedral space between two NiO$_6$ octahedra that cant toward one another. Figures made using VESTA.}
\label{fig:H}
\end{figure}

\begin{figure}[!h]
\center
\includegraphics[width=\textwidth]{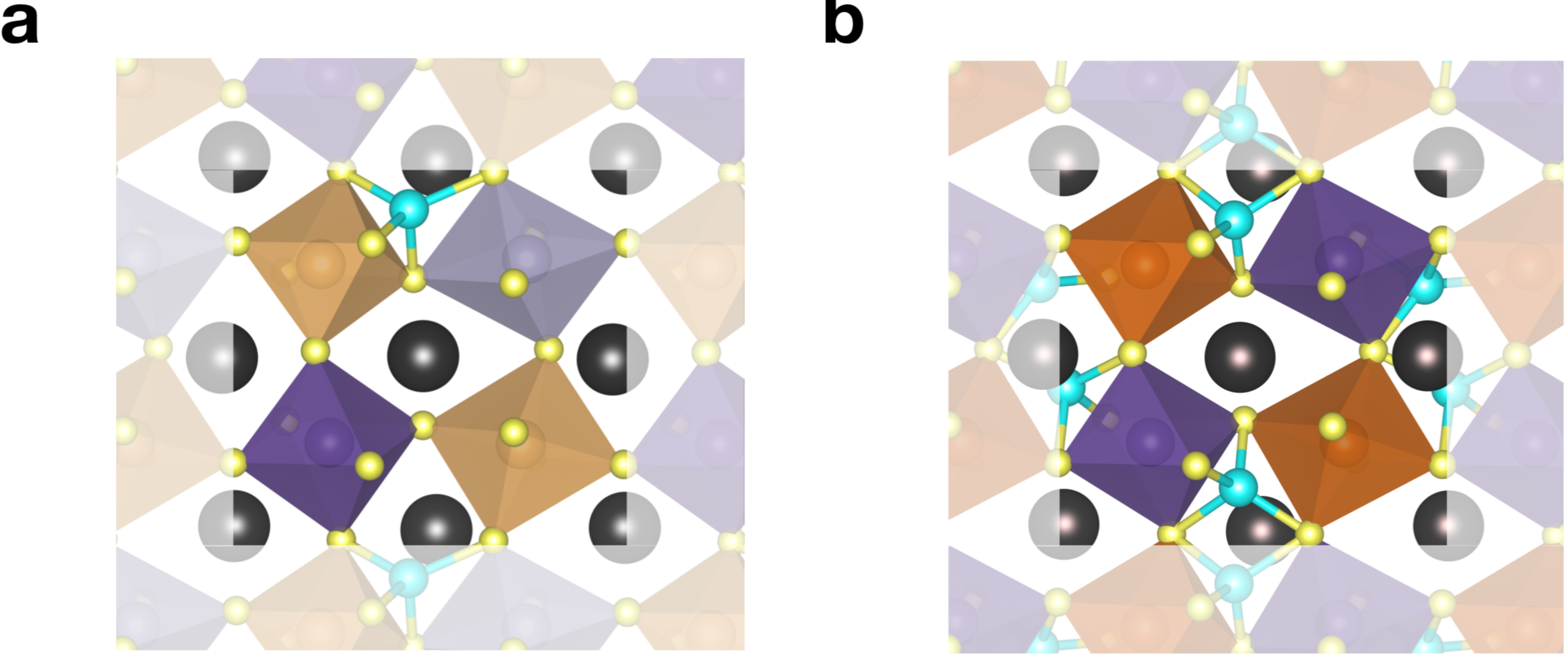}
\caption{One $\sqrt{2}\times\sqrt{2}\times 2$ cell for the (a) low energy structure for Li$_{\frac{1}{4}}$SNO; and, (b) the low energy structure for LiSNO. Sm shown in dark gray; Ni in purple and orange, where the lighter hues indicate Ni$^{3+}$ and the darker hues Ni$^{2+}$; O in yellow; and, Li in cyan. The lithium are tetrahedrally coordinated by oxygen in the tetrahedral space between two NiO$_6$ octahedra that cant toward one another. Figures made using VESTA.}
\label{fig:Li}
\end{figure}

%\section{Comparison of implicit (added electrons) and explicit (added interstitials) electron doping}
To explicitly electron dope SNO, we include neutral hydrogen and lithium into our calculations. In both cases the valence electron localized on a nearby NiO$_6$ octahedron. The structures for H$_{\frac{1}{4}}$SNO and HSNO are shown in Fig.~\ref{fig:H} and Li$_{\frac{1}{4}}$SNO and LiSNO are shown in Fig.~\ref{fig:Li}.  For further details of the inclusion of hydrogen and lithium and the resulting low energy structures see the works of Zhang et al.~\cite{Zhang2018} and Sun et al.~\cite{Sun2018}, respectively. When the lattice parameter in the (110) orthorhombic direction is allowed to relax (following the epitaxial constraints in the experimental thins films), we observe a decrease in the band gap and a splitting of the states in the conduction band due to the distortions of the NiO$_6$ octahedra (see Fig.~\ref{fig:inter}). These large distortions are due to creating a local binding environment for the interstitial ions. For example see the structure of Li$_{0.25}$SmNiO$_3$ Fig~\ref{fig:Li}(a): the NiO$_6$ closest to the added lithium does not expand so that the tetrahedral coordination is preserved and the lithium valence electron localizes on a neighboring NiO$_6$ octahedron causing that next-nearest neighboring NiO$_6$ octahedron to expand. In the case of LiSmNiO$_3$ (Fig~\ref{fig:Li}(b)), all the  NiO$_6$ octahedra have expanded; however, they have done so in a non-uniform manner to preserve the tetrahedral coordination of the lithium.

\begin{figure}[!h]
\center
\includegraphics[width=1.\textwidth]{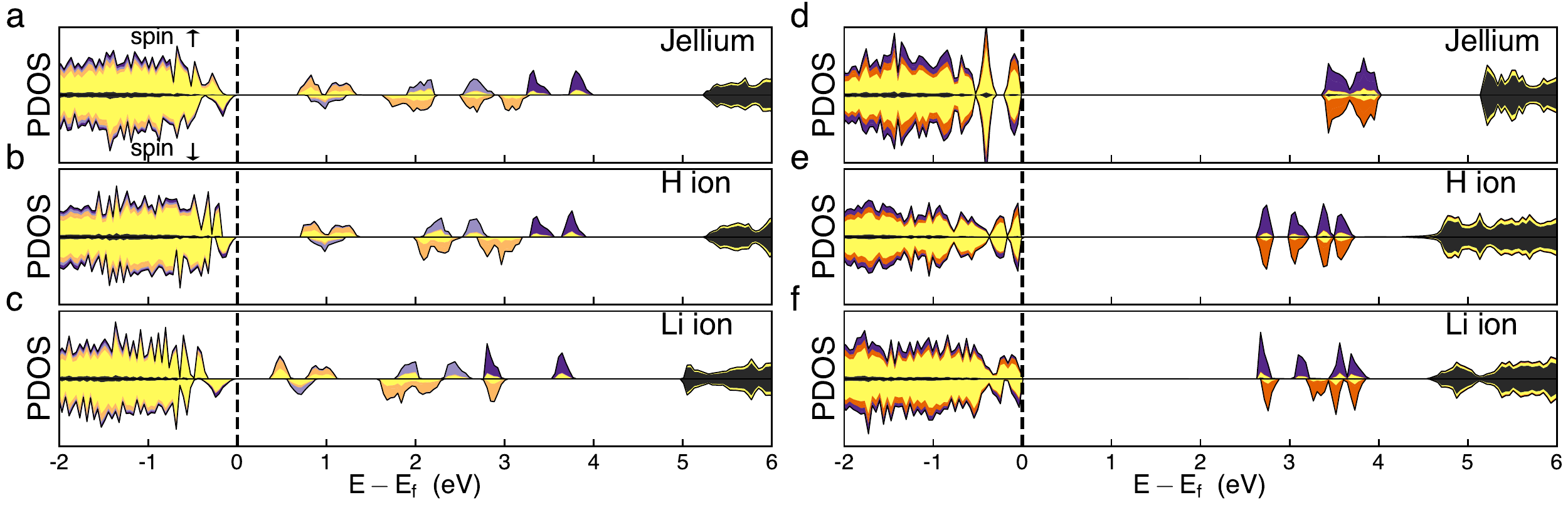}
\caption{Spin-polarized PDOS for electron-doped SNO with (a) 1/4 added e$^-$/Ni with a positive compensating background charge; (b) 1/4  H/Ni; (c) 1/4  Li/Ni;  (d) 1 added e$^-$/Ni with a positive compensating background charge; (e) 1  H/Ni; and, (f) 1 Li/Ni. The element specific PDOS is shown in yellow for O, orange and purple for Ni with a positive or negative magnetic moment, respectively, and dark gray for Sm. \niii~is displayed in the lighter hues and \nii~in the darker hues. In panel (a) and (d) the volume is restricted to that of undoped SNO. For panels (b), (c), (e), and (f) the lattice parameters have been relaxed in the (110) $Pbnm$ direction (experimental epitaxial constraint).}
\label{fig:inter}
\end{figure}

\end{document}